\documentclass[lettersize,journal]{IEEEtran}
\usepackage{amsmath,amsfonts}
\usepackage{algorithmic}
\usepackage{algorithm}
\usepackage{array}
\usepackage[caption=false,font=footnotesize,labelfont=rm,textfont=rm]{subfig}
\usepackage{textcomp}
\usepackage{stfloats}
\usepackage{url}
\usepackage{verbatim}
\usepackage{graphicx}
\usepackage{cite}
\usepackage{orcidlink}
\usepackage{tikz}
\usepackage{hyperref}
\usepackage{xcolor}
\begin{document}

\title{PAPR Reduction with Pre-chirp Selection for Affine Frequency Division Multiplexing\\
\thanks{The corresponding author is Yin Xu (e-mail: xuyin@sjtu.edu.cn).}}

\author{Haozhi Yuan\hspace{-1.5mm}$^{~\orcidlink{0009-0001-7761-5447}}$, Yin Xu\hspace{-1.5mm}$^{~\orcidlink{0000-0003-4429-5982}}$,~\IEEEmembership{Senior Member,~IEEE}, Xinghao Guo\hspace{-1.5mm}$^{~\orcidlink{0009-0003-5584-4024}}$, Yao Ge\hspace{-1.5mm}$^{~\orcidlink{0000-0002-3293-2051}}$,~\IEEEmembership{Member,~IEEE}, Tianyao Ma\hspace{-1.5mm}$^{~\orcidlink{0009-0004-1114-308X}}$,\\
Haoyang Li\hspace{-1.5mm}$^{~\orcidlink{0000-0002-4816-4325}}$,~\IEEEmembership{Member,~IEEE}, 
Dazhi He\hspace{-1.5mm}$^{~\orcidlink{0000-0001-6637-0088}}$,~\IEEEmembership{Senior Member,~IEEE}, Wenjun Zhang\hspace{-1.5mm}$^{~\orcidlink{0000-0001-8799-1182}}$,~\IEEEmembership{Fellow,~IEEE}}

\markboth{}%
{Shell \MakeLowercase{\textit{et al.}}: A Sample Article Using IEEEtran.cls for IEEE Journals}


\maketitle

\begin{abstract}
Affine frequency division multiplexing (AFDM) is a promising new multicarrier technique for high-mobility communications based on discrete affine Fourier transform (DAFT). By properly tuning two parameters in the DAFT module, the effective channel in the DAFT domain can completely circumvent path overlap, thereby constituting a full representation of delay-Doppler profile. However, AFDM has a crucial problem of high peak-to-average power ratio (PAPR), stemming from randomness of modulated symbols. To reduce the PAPR of AFDM, a novel algorithm named grouped pre-chirp selection (GPS) is proposed in this letter. The GPS varying the pre-chirp parameter across subcarrier groups in a non-enumerated manner and then selects the signal with the smallest PAPR among all candidate signals. We detail the operational procedures of the GPS algorithm, analyzing GPS from four aspects: PAPR performance, computational complexity, spectral efficiency, and bit error rate (BER) performance. Simulations indicate the effectiveness of the proposed GPS in reducing PAPR. At the cost of slight communication performance, AFDM with GPS can achieve better PAPR performance than orthogonal time frequency space (OTFS) while maintaining a lower modulation complexity.



\end{abstract}


\section{Introduction}
\IEEEPARstart{T}{he} next generation wireless systems (beyond 5G/6G) is anticipated to accommodate high-mobility scenarios characterized by high relative velocities, which result in substantial Doppler frequency shifts and cause the wireless channel to exhibit rapidly time-varying characteristics. In existing 4G/5G, the prevalent multicarrier technique, i.e., orthogonal frequency division multiplexing (OFDM), exhibits limited robustness against carrier frequency offset (CFO) \cite{CFO}, thus the performance degrades significantly in high-mobility scenarios.


In this context, a series of advanced waveform designs are proposed. Among these, affine frequency division multiplexing (AFDM) stands out as a promising multicarrier modulation scheme utilizing the discrete affine Fourier transform (DAFT) \cite{AFDM1,AFDM2}. By appropriately tuning two DAFT chirp parameters, here called the pre-chirp parameter and the post-chirp parameter, subchannels originating from different propagation paths can be separated. This leads to a sparse and quasi-static representation of the effective channel, thereby enabling AFDM to attain full diversity order in doubly dispersive channels. Moreover, DAFT serves as a generalized form of the discrete Fourier transform (DFT), ensuring AFDM to remain fully compatible with OFDM. Compared to orthogonal time frequency space (OTFS) \cite{OTFS}, another emerging multicarrier technique for high-mobility communications, AFDM requires less pilot guard overhead and has lower modulation complexity \cite{OTFS_2_AFDM}. AFDM has garnered increasing attention in the literature due to its advantages in high-mobility communications and compatibility. Recent studies have focused on various aspects of AFDM such as channel estimation \cite{Ch_Es}, equalization \cite{EQ}, and index modulation (IM) \cite{IM,pre-chirp-domain}.


Despite the advantages, AFDM encounters a critical challenge of high peak-to-average power ratio (PAPR), which is theoretically identical to that in OFDM. In OFDM, there are several methods for PAPR reduction, such as coding \cite{coding}, selective mapping (SLM) \cite{SLM}, Tone Reservation (TR) \cite{TR}. On the contrary, researches on PAPR reduction in AFDM rarely exist. A DAFT-spread AFDMA scheme is proposed in \cite{DAFT_spread}. However, this method is primarily focused on multiple access (MA), making it less suitable for scenarios with a small number of users, as it effectively results in single-carrier transmission for each user. 



Inspired by \cite{pre-chirp-domain}, where a distinct pre-chirp parameter is assigned to each subcarrier to enhance transmission efficiency, an opportunity arises to reduce the PAPR by substituting the original uniform pre-chirp parameter. In this letter, a grouped pre-chirp selection (GPS) algorithm is proposed to reduce PAPR in AFDM, which solely modifies the parameters within the pre-chirp module. First, an optimal algorithm is proposed that exhaustively enumerates all feasible pre-chirp values across every subcarrier. Subsequently, to reduce the complexity, the optimal algorithm has been improved from three aspects, and the modified algorithm is referred to as GPS.





\section{System Model}
\subsection{Conventional AFDM}
\begin{figure}[h]
	\centering
	\includegraphics[width=8.5cm]{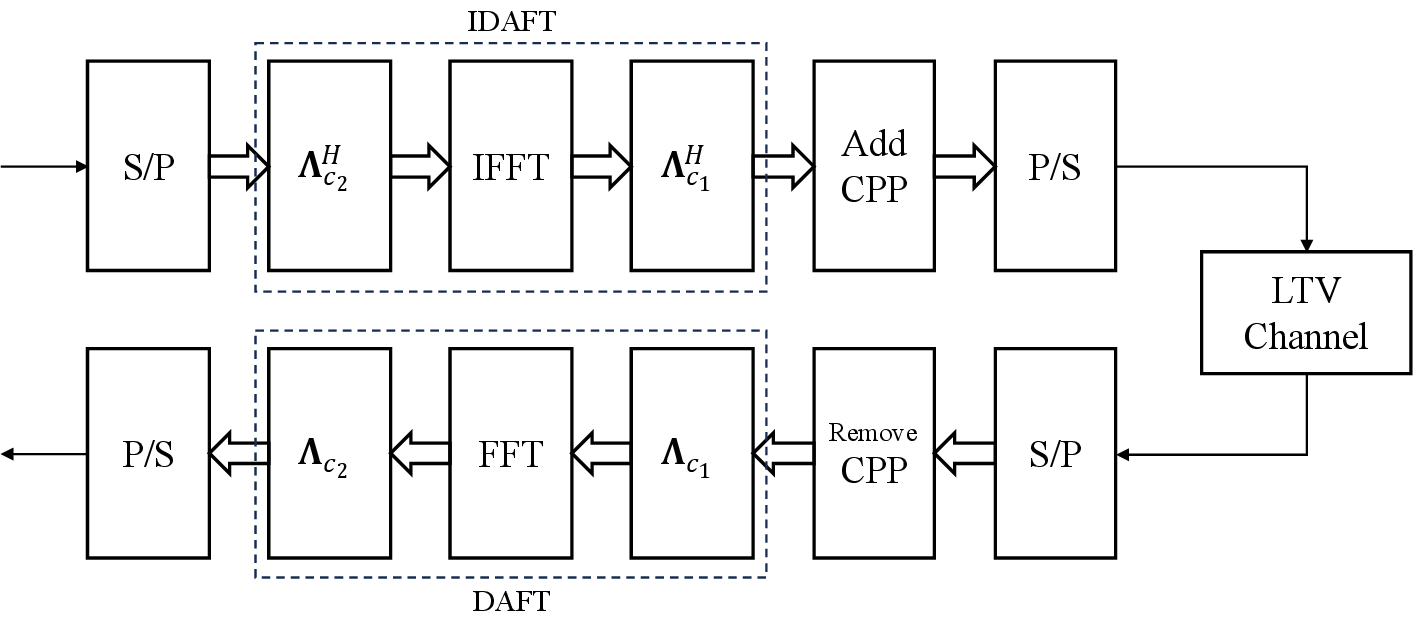}
	\caption{AFDM block diagram.}
	\label{Fig_1}
\end{figure}
The conventional AFDM block diagram is depicted in Fig. \ref{Fig_1}. Data symbols are initially placed in the DAFT domain. At the transmitter, inverse DAFT (IDAFT) is employed to map the symbols from the DAFT domain into the time domain. With $N$ chirp subcarriers, the IDAFT operation can be formulated as
\begin{equation}
s[n]=\frac{1}{\sqrt{N}}\sum_{m=0}^{N-1}x[m]\cdot e^{j2\pi\left(c_1n^2+c_2m^2+mn/N\right)},
\end{equation}
where $x[m]$ denotes the modulated symbol in the DAFT domain, and $s[n]$ signifies the signal in the time domain. According to (1), the frequencies of all subcarriers exhibit periodic and linear variations over time in AFDM. After adding chirp-periodic prefix (CPP), the time-domain signal $s[n]$ undergoes propagation through the linear time-varying (LTV) channel whose impulse response can be represented as
\begin{equation}
h_n[l]=\sum\limits_{p=1}^Ph_pe^{-j2\pi f_pn}\delta(l-l_p),
\end{equation}
where $P$ is the number of paths, $\delta(\cdot)$ is the Dirac delta function, $h_p$, $f_p$, and $l_p$ are the complex gain, Doppler shift, and the integer delay associated with the $p$-th path, respectively.

The DAFT/IDAFT operations and the effect of the LTV channel can both be expressed in matrix form. The input-output relation in the time domain, i.e., the effect of the LTV channel at the receiver, can be formulated as
\begin{equation}
\mathbf{r}=\mathbf{H}\mathbf{s}+\mathbf{w},
\end{equation}
where $\mathbf{s}$, $\mathbf{r}\in\mathbb{C}^{N\times1}$, $\mathbf{w}\sim\mathcal{CN}\left(\mathbf{0},N_{0}\mathbf{I}\right)\in\mathbb{C}^{N\times1}$, and 
\begin{equation}
\mathbf{H}=\sum_{p=1}^{P}h_{p}\mathbf{\Gamma}_{{\mathrm{CPP}_{p}}}\mathbf{W}^{{f_{p}}}\boldsymbol{\Pi}^{{l_{p}}},
\end{equation}
where $\mathbf{\Pi}$ is the permutation matrix,
$\mathbf{W}$ is the diagonal matrix
\begin{equation}
\mathbf{W}=\mathrm{diag}(e^{-j2\pi n},n=0,1, \ldots, N-1),
\end{equation}
and $\boldsymbol{\Gamma}_{\mathrm{CPP}_{p}}$ is an $N\times N$ diagonal matrix
\begin{equation}
\begin{aligned}
&\boldsymbol{\Gamma}_{\mathrm{CPP}_p} =\\
&\operatorname{diag}\left(\!\left\{\!\!\begin{array}{ll}
e^{-j 2 \pi c_1\left(N^2-2 N\left(l_p-n\right)\right)} \!& n<l_p \\
1 \!& n \geq l_p
\end{array}\right.\!\!\!\!,
n\!=\!0, \ldots, N-1\!\right) .
\end{aligned}
\end{equation}

The DAFT matrix is expressed as
\begin{equation}
\mathbf{A}=\mathbf{\Lambda}_{{c_{2}}}\mathbf{F}\mathbf{\Lambda}_{{c_{1}}},
\end{equation}
where $\mathbf{F}\in\mathbb{C}^{N\times N}$ is the normalized $N$-point DFT matrix and
\begin{equation}
\boldsymbol{\Lambda}_{c_i}=\operatorname{diag}(e^{-j2\pi c_i n^2},n=0,1, \ldots ,N-1,i=1,2).
\end{equation}
It is obvious that $\mathbf{A}$ is unitary. Consequently, the IDAFT matrix can be expressed as
\begin{equation}
\mathbf{A}^{-1}=\mathbf{A}^H=\mathbf{\Lambda}_{c_1}^H\mathbf{F}^H\mathbf{\Lambda}_{c_2}^H.
\end{equation}

Let $\mathbf{x}=[x[0],x[1],...,x[N-1]]^T\in\mathbb{C}^{N\times1}$ denotes the symbols in the DAFT domain at the transmitter. Similarly, let $\mathbf{y}=[y[0],y[1],...,y[N-1]]^T\in\mathbb{C}^{N\times1}$ denotes the received signal in the DAFT domain. The comprehensive input-output relation in the DAFT domain can thus be formulated as
\begin{equation} 
{\mathbf y} \!=\! \mathbf {A}\mathbf {r} \!=\! \sum _{p=1}^{P} h_{p} \mathbf {A}{\boldsymbol{\Gamma }}_{\mathrm {CPP}_{p}} {\mathbf{W}^{f_p}} {\boldsymbol{\Pi }}^{l_{p}} \mathbf {A}^{H}{\mathbf x}+ \mathbf {A}{\mathbf w} = {\mathbf H}_{\mathrm {eff}} {\mathbf x}+ \widetilde {\mathbf w},
\end{equation}
where $\widetilde {\mathbf{w}}\sim\mathcal{CN}\left(\mathbf{0},N_{0}\mathbf{I}\right)\in\mathbb{C}^{N\times1}$ still holds due to $\mathbf {A}$ being a unitary matrix, and
\begin{equation} 
{\mathbf H}_{\mathrm {eff}}=\sum _{p=1}^{P} h_{p} \mathbf {A}{\boldsymbol{\Gamma }}_{\mathrm {CPP}_{p}} {\mathbf{W}^{f_p}} {\boldsymbol{\Pi }}^{l_{p}} \mathbf {A}^{H}.
\end{equation}

According to \cite{AFDM1}, achieving full diversity in LTV channels with integer Doppler shift requires setting the pre-chirp parameter $c_{2}$ to any irrational value. Additionally, the minimum permissible value for the post-chirp parameter $c_{1}$ is given by
\begin{equation}
c_1=\frac{2\alpha_{\max}+1}{2N},
\end{equation}
where $\alpha_{\max}$ is the integer part of the maximum Doppler shift normalized with respect to the subcarrier spacing. The value of $c_1$ is crucial for AFDM to achieve full diversity, whereas the constraint on $c_2$ is loose, which provides sufficient flexibility to reduce PAPR by optimizing $c_2$.

\subsection{AFDM with GPS}
\begin{figure}[h]
	\centering
	\includegraphics[width=8.85cm]{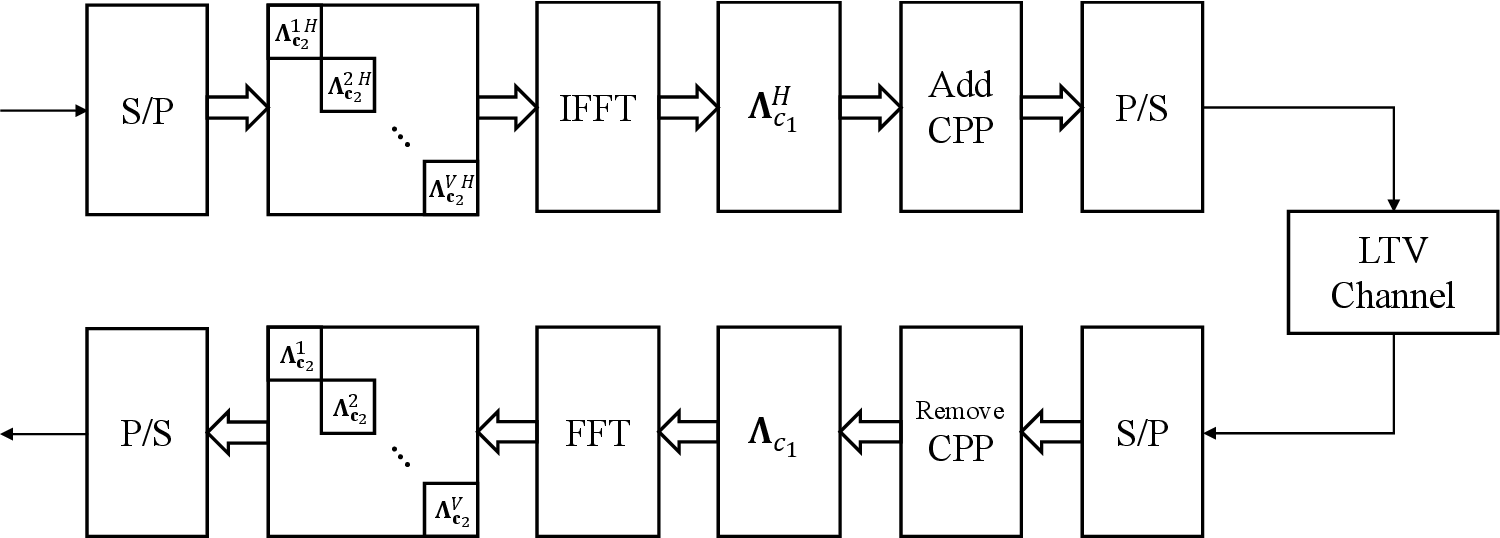}
	\caption{AFDM with GPS block diagram.}
	\label{Fig_2}
\end{figure}
The block diagram depicted in Fig. \ref{Fig_2} illustrates AFDM with GPS, where the pre-chirp module is divided into $V$ groups, each containing $M=\frac{N}{V}$ pre-chirp subcarriers. Consequently, the subdiagonal matrix of each group can be expressed as
\begin{equation}
\mathbf{\Lambda}_{c_2}^v=\operatorname{diag}(e^{-j2\pi c_{2,m} m^2},m=(v-1)M,\ldots,vM-1),
\end{equation}
where $v=1,\ldots,V$. Distinct $c_2$ values are employed across each subcarrier, where $c_{2,m}\in\Omega_m$ denotes the value of $c_2$ on the $m$-th subcarrier, and $\Omega_m$ represents the set of feasible values for $c_{2,m}$. It should be emphasized that within the $v$-th group, all $c_{2,m}$ share a common element index $w_v$ in $\Omega_m$, rather than the same specific value. The resultant time-domain signals can be mathematically represented as
\begin{align}
s[n]=\frac{1}{\sqrt{N}}\sum_{m=0}^{N-1}x[m]\cdot e^{j2\pi\left(c_1n^2+c_{2,m}m^2+mn/N\right)}.
\end{align}
Such a modification is intended to reduce the PAPR. The orthogonality between any pair of subcarriers remains unaffected according to \cite{pre-chirp-domain}. Furthermore, the sparse property of the effective channel ${\mathbf H}_{\mathrm {eff}}$ persists. Given that both $\mathbf{\Lambda}_{{c_{2}}}^H$ and $\mathbf{\Lambda}_{{c_{2}}}$ are diagonal matrices, altering the values of the diagonal elements does not influence the locations of the non-zero elements. The persistence of the above properties forms the basis for our proposed GPS algorithm.

\section{Proposed PAPR Reduction Scheme}
In this section, we elaborate on two proposed schemes for PAPR reduction: the optimal algorithm and the GPS algorithm.

Within each block, the PAPR is formulated as
\begin{equation}
\mathrm{PAPR}=\frac{P_{\mathrm{max}}}{P_{\mathrm{avg}}}=\frac{\underset{n=0,\ldots,N-1}\max( |s[n]|^{2})}{\frac{1}{N}\sum_{n=0}^{N-1}|s[n]|^{2}},
\end{equation}
where $s[n]$ is the $n$-th sample of the time-domain signals. Although PAPR is defined for continuous-time bandpass signals, with no less than 4-fold oversampling, $s[n]$ and the continuous-time bandpass signals have almost the same PAPR \cite{oversampling}. To quantitatively characterize the PAPR metric, the complementary cumulative distribution function (CCDF) is utilized, which is defined as
\begin{equation}
\mathrm{CCDF}=\mathrm{Pr}(\mathrm{PAPR}>\mathrm{PAPR_0}),
\end{equation}
where $\mathrm{PAPR_0}$ represents a specific threshold value of the PAPR, and $\operatorname{Pr}(\cdot)$ denotes the probability function.

An optimal algorithm is firstly proposed, which can be described as follows. Initially, $N-1$ finite sets $\Omega_m$ $(m=1,\dots,N-1)$ are created with irrational elements to satisfy the value of $c_2$. Each $\Omega_m$ is assumed to contain $W$ elements and $\Omega_m(w)$ denotes the $w$-th element. By exhaustively enumerating all $W$ elements of $\Omega_m$ across every subcarrier, the optimal PAPR can theoretically be achieved. However, this approach incurs prohibitively high computational costs because each PAPR calculation involves an IDAFT matrix multiplication, whose complexity is normalized to $\mathcal{O}\left(1\right)$. Consequently, a total of $\mathcal{O}\left(W^{N-1}\right)$ computational complexity are required to obtain the final result. Note that the first subcarrier (where $m=0$) requires no alteration as the initial phase remains constant at $0$ according to (14), which corresponds to the exponent $N-1$.

\begin{algorithm}[]
    \caption{GPS with $W=2$ in AFDM System for PAPR Reduction}
	\label{algorithm1} 
	\renewcommand{\algorithmicrequire}{\textbf{Input:}}
	\renewcommand{\algorithmicensure}{\textbf{Output:}}
	\begin{algorithmic}[1]
		\REQUIRE Number of subcarriers $N$, number of groups $V$, number of subcarriers per group $M$, post-chirp matrix $\mathbf{\Lambda}_{c_1}$, normalized FFT matrix $\mathbf{F}$, symbol vector ${\mathbf{x}}$, feasible values set $\mathbf{\Omega}_m\in\mathbb{C}^{N\times2}$;
            \STATE Initialize $\mathbf{\Lambda}_{c_2}^{(0)}$ and $\mathrm{PAPR}^{(0)}$ as \hspace{0.5mm}(23) \hspace{0.5mm}and \hspace{0.5mm}(24), respectively;
            \STATE Initialize $\mathbf{c}_{2,m}^{(0)}=\mathbf{\Omega}_m(:,1)$;
            \STATE $\mathrm{PAPR}_{\min}=\mathrm{PAPR}^{(0)}$;
            \FOR{$l=1:V$}
                \STATE $\mathbf{c}_{2,m}^{(l)}\left((l\!-\!1)M\!+\!1\!:\!lM\right)\!=\!\mathbf{\Omega}_m\left((l\!-\!1)M\!+\!1\!:\!lM,2\right)$;
                \STATE Calculate $\mathbf{\Lambda}_{c_2}^{(l)}$ using $\mathbf{c}_{2,m}^{(l)}$;
                \STATE Calculate $\mathrm{PAPR}^{(l)}$ according to (24);
                \IF{$\mathrm{PAPR}^{(l)}\geq\mathrm{PAPR}_{\min}$}
                    \STATE $\mathbf{c}_{2,m}^{(l)}\left((l\!-\!1)M\!+\!1\!:\!lM\right)\!=\!\mathbf{\Omega}_m\left((l\!-\!1)M\!+\!1\!:\!lM,1\right)$;
                \ELSE
                    
                    \STATE $\mathrm{PAPR}_{\min}=\mathrm{PAPR}^{(l)}$;
                \ENDIF
            \ENDFOR
            \STATE Calculate $\mathbf{\Lambda}_{c_2}^*$ using $\mathbf{c}_{2,m}^{(V)}$;
		
		\ENSURE $\mathrm{PAPR}_{\min}$ and $\mathbf{\Lambda}_{c_2}^*$.
	\end{algorithmic} 
\end{algorithm}

Given the impractically high computational complexity of the aforementioned scheme, we further propose a GPS algorithm to achieve a balanced trade-off between performance and computational efficiency. Compared to the optimal algorithm, the GPS greatly reduces the computational complexity with no significant performance degradation. 

In general, the GPS algorithm tunes the values of $c_{2,m}$ on each subcarrier by group in a non-enumerated manner and then selects the signal with the smallest PAPR among all the candidate signals. On the basis of the optimal algorithm, the additional design of the GPS algorithm can be summarized in the following three aspects.




\subsubsection{$\Omega_m$}
The principle governing the selection of elements in $\Omega_m$ is considered. According to (14), the phase of each term in the summation process is given by $\mathrm{arg}(x[m])+2\pi(c_1 n^2+c_{2,m}m^2+mn/N)$. Within each block, tuning the value of $c_{2,m}$ does not affect average power but impacts peak power. For future convenience, we deviate from the notation of the peak power employed in (15) by denoting it as $|s|_{\max}^2$. For a large number of subcarriers $N$, according to the central limit theorem (CLT), the envelope of the time-domain signal approximately obeys a Gaussian distribution $s[n]\sim\mathcal{CN}\left(0,1\right)$, where $1$ is the normalized transmit power. Therefore, the cumulative distribution function (CDF) of $|s[n]|^2$ is $F_{|s[n]|^2}(\gamma)=1-e^{-\gamma}$ $(\gamma \geq 0)$. Based on \cite{PAPR}, the CDF of $|s|_{\max}^2$ can be approximated as
\begin{equation}
F_{|s|_{\max}^2}(\gamma)=\exp \left(-e^{-\gamma} N \sqrt{\frac{\pi}{3} \ln{N}}\right).
\end{equation}
Varying $c_{2,m}$ on any one subcarrier can generate $W$ candidate signals whose peak power is respectively denoted as $|s|_{\max,w}^2$, $w=1,\dots,W$. Obviously, there is $\mathrm{Pr}(|s|_{\max,w}^2>\gamma,\forall w) < \mathrm{Pr}(|s|_{\max,1}^2>\gamma)$, which implies that increasing $W$ improves the PAPR performance. On the other hand, values of elements in $\Omega_m$ also affect the PAPR performance. To be specific, given a fixed $W$, in order to minimize $\mathrm{Pr}(|s|_{\max,w}^2>\gamma,\forall w)$, the time-domain signals corresponding to different elements in $\Omega_m$ should be as uncorrelated as possible. The scenario where $W=2$ and only $c_{2,m_0}$ is tuned is considered first for simplicity. The correlation coefficient of the two corresponding candidate signals is expressed as
\begin{equation}
\begin{aligned}
\rho & =\frac{\operatorname{Cov}\left(s[n]_1, s[n]_2\right)}{\sqrt{D\left(s[n]_1\right) D\left(s[n]_2\right)}} \\ & =E\left(s[n]_1 s[n]_2^*\right) \\ & =\frac{1}{N}E\!\!\left(|x[m_0]|^2 e^{j 2 \pi (\Omega_{m_0}(1)-\Omega_{m_0}(2))m_0^2}+\!\!\!\sum_{m \neq m_0}|x[m]|^2\right) \\ & =\frac{N-1+e^{j 2 \pi (\Omega_{m_0}(1)-\Omega_{m_0}(2))m_0^2}}{N},
\end{aligned}
\end{equation}
where $s[n]_w$ denotes the signal generated by $c_{2,m_0}=\Omega_{m_0}(w)$. To minimize $|\rho|$, it should satisfy 
\begin{equation}
2 \pi (\Omega_{m_0}(1)-\Omega_{m_0}(2))m_0^2=\pi.
\end{equation}
This principle can be generalized to every subcarrier, thus values of elements in $\Omega_m$ are given as
\begin{equation}
\Omega_m (w) = (-1)^{w-1}\frac{1}{4m^2},
\end{equation}
where $w=1,2$ corresponds to the two elements. However, given that the pre-chirp value should be irrational \cite{AFDM1}, (20) is accordingly substituted with
\begin{equation}
\Omega_m (w)=\begin{pmatrix}-1\end{pmatrix}^{w-1}\frac{\pi\cdot10^k}{4m^2 \lfloor\pi \cdot 10^k\rfloor},
\end{equation}
where $\lfloor\cdot\rfloor$ denotes the floor function, and $k$ determines the precision of decimals retained. This principle can be further generalized to cases where $\Omega_m$ contains $W$ ($W\geq2$) elements. The $W$ feasible values is given as
\begin{equation}
\Omega_m (w)=\frac{(w-\frac{1}{2})\pi\cdot10^k}{Wm^2 \lfloor\pi \cdot 10^k\rfloor},
\end{equation}
where $w=1,\cdots,W$. Through (22), it can be observed that the diagonal elements of the pre-chirp module are uniformly distributed over the unit circle. However, as $W$ increases, the phase difference between the elements inevitably diminishes. Consequently, the performance enhancement gradually slows down while the complexity escalates linearly. Therefore, to achieve a good balance between performance and complexity, employing two or three elements in each $\Omega_m$ is sufficient, as substantiated by simulations.

\subsubsection{Non-enumerated algorithm}
Adopting a non-enumerated algorithm with fewer iterations can substantially mitigate complexity. Specifically, the pre-chirp values in $\Omega_m$ are sequentially selected across all subcarriers, rather than enumerating all possible permutations. Each pre-chirp adjustment that reduces PAPR will permanently replace the prior value; otherwise, it is reverted. Precisely, when $W=2$, the computational complexity of the enumerated algorithm and the non-enumerated algorithm is $\mathcal{O}\left(2^{N-1}\right)$ and $\mathcal{O}\left(N-1\right)$, respectively.

\subsubsection{Grouping}
Complexity can be further alleviated by grouping subcarriers. The number of subcarriers included in any one group is arbitrary in principle, but each group is set with equal size for convenience. By dividing the subcarriers into $V$ groups, the computational complexity is further reduced from $\mathcal{O}\left(N-1\right)$ to $\mathcal{O}\left(V\right)$ per block. In addition to the number of groups, the grouping pattern also affects the performance. Two grouping patterns are examined: adjacent grouping (where adjacent subcarriers are grouped together) and comb grouping (where spaced subcarriers are grouped together), which will be shown in the subsequent section.


The proposed GPS with $W=2$ is outlined in Algorithm \ref{algorithm1} as an example. The initialization step involves setting the pre-chirp value on each subcarrier to a positive number. Here, we set $c_{2,m}=\frac{\pi}{12.56m^2}$ where $k=2$ on each subcarrier. Consequently, the initialized pre-chirp matrix is given as
\begin{equation}
\mathbf{\Lambda}_{c_2}^{(0)}=\operatorname{diag}\left(1,e^{j\frac{2\pi^2}{12.56}},\ldots,e^{j\frac{2\pi^2}{12.56}}\right)_{N\times N}.
\end{equation}
The corresponding initial PAPR is subsequently computed as 
\begin{equation}
\mathrm{PAPR}^{(0)}=\frac{\max(|\mathbf{\Lambda}_{c_1}^H\mathbf{F}^H\mathbf{\Lambda}_{c_2}^{H(0)}\mathbf{x}|^2)}{\mathrm{mean}(|\mathbf{\Lambda}_{c_1}^H\mathbf{F}^H\mathbf{\Lambda}_{c_2}^{H(0)}\mathbf{x}|^2)}.
\end{equation}
During each iteration, the pre-chirp values within a designated group undergo complete reversal. Then, a new $\mathrm{PAPR}^{(l)}$ is calculated using an updated pre-chirp matrix $\mathbf{\Lambda}_{c2}^{(l)}$. Upon completion of the algorithm, the minimum PAPR among $(V+1)$ values, denoted as $\mathrm{PAPR}_{\min}$, along with its corresponding pre-chirp matrix $\mathbf{\Lambda}_{c_2}^*$, is obtained as the output.

\section{Numerical Results}
An AFDM system with 16QAM under LTV channel is considered. Without loss of generality, channel coefficients $h_{i}\sim\mathcal{CN}\left(0,1\right)$ are generated. The number of subcarriers is set as $N=64$. At the receiver, the channel information is assumed to be perfectly known. Totally, $\lceil{\log _2(W^V)}\rceil$ extra side bits are needed for decoding, and the minimum mean square error (MMSE) equalization is applied.
\begin{figure}[htbp]
    \centering
	\subfloat[Different $W$ and $\Omega_m$\label{Fig_3a}]{
	\includegraphics[width=.47\columnwidth]{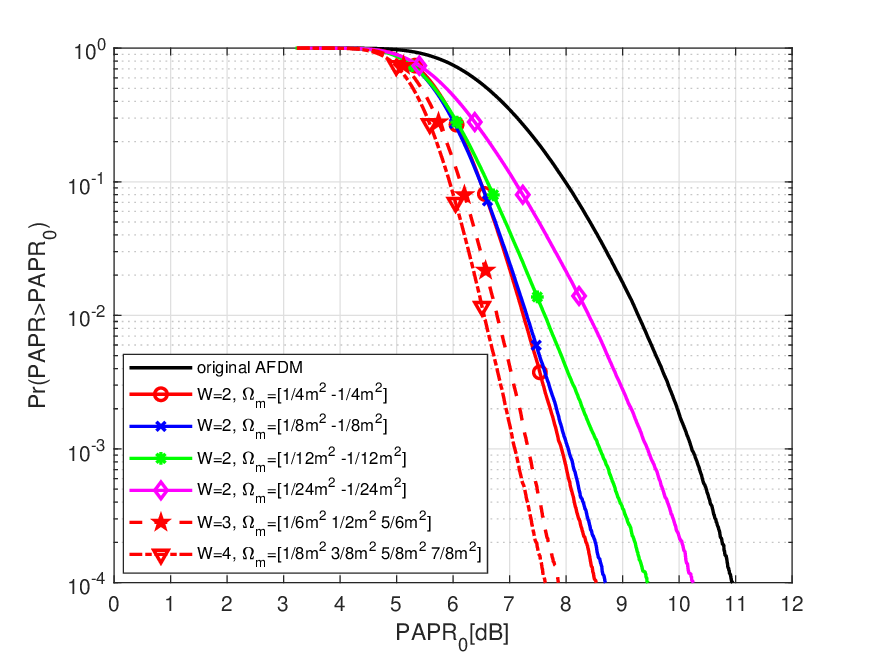}
    }
	\subfloat[Different $V$ and grouping patterns\label{Fig_3b}]{
	\includegraphics[width=.47\columnwidth]{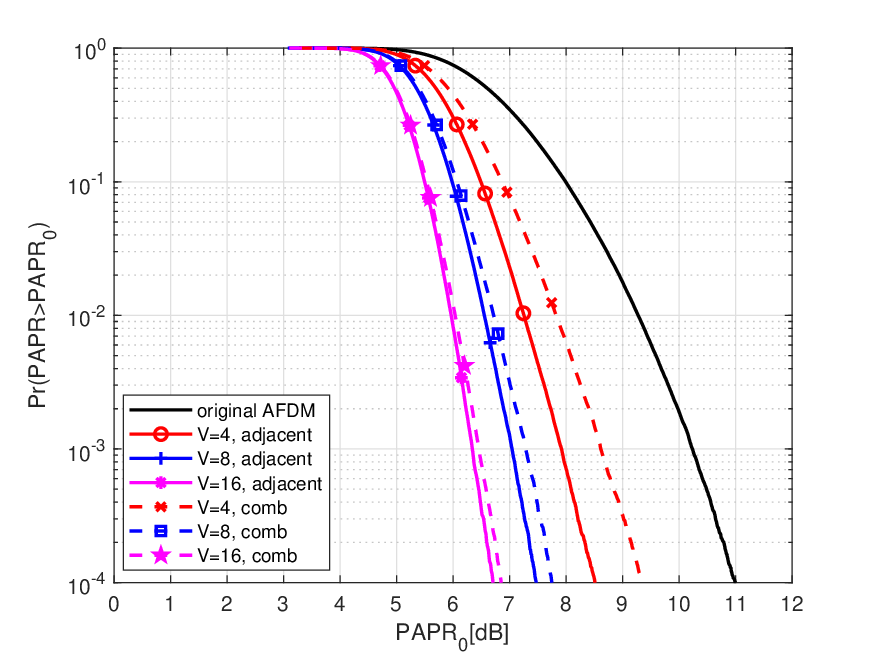}
	}
	\caption{PAPR performance of 16QAM AFDM with GPS.}
	\label{Fig_3}
\end{figure}

In Fig. \ref{Fig_3a}, simulation results of GPS are presented for different values and number of elements in $\Omega_m$, where $V=4$ and adjacent grouping pattern are applied. In terms of values of elements, the results demonstrate that a larger phase difference yields improved performance, assuming an equivalent number of elements, which aligns with the conclusion derived from our derivation. In terms of number of elements $W$, the performance gain from the increase of number converges quickly. 

Fig. \ref{Fig_3b} illustrates the impact of the number of groups and different grouping strategies on performance, where $W=2$ and $\Omega_m=[\frac{1}{4m^2}, -\frac{1}{4m^2}]$ are employed. It can be observed that adjacent grouping consistently outperforms comb grouping. Targeting the CCDF at a level of $10^{-4}$, the adjacent grouping pattern demonstrates performance improvements of approximately $0.7$ dB, $0.3$ dB, and $0.1$ dB over the comb grouping pattern for configurations with $4$, $8$, $16$ groups, respectively. In addition, increasing the number of groups could enhance the effectiveness of PAPR reduction. However, the rate of performance improvement diminishes as the number of groups increases, which indicates a significant increase in complexity. To balance complexity and performance, it is advisable to keep the number of groups small, provided that the desired PAPR reduction threshold is achieved.

\begin{figure}[h]
	\centering
	\includegraphics[width=5.8cm]{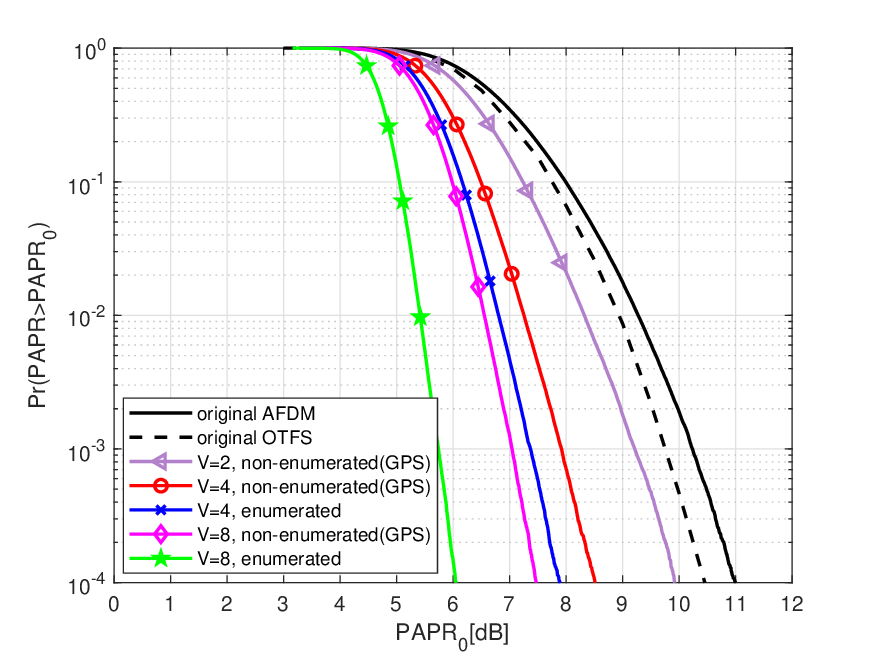}
	\caption{PAPR performance of 16QAM AFDM with GPS, enumerated algorithm, and OTFS.}
	\label{Fig_4}
\end{figure}

In Fig. \ref{Fig_4}, the performance disparity between the enumerated method and the GPS is depicted, where $W=2$, $\Omega_m=[\frac{1}{4m^2}, -\frac{1}{4m^2}]$, and the adjacent grouping pattern are applied. The enumerated method consistently outperforms the GPS. However, this enhancement comes at the cost of significantly increased computational complexity. Notably, the GPS with $V=8$ ($\mathcal{O}\left(8\right)$) achieves a performance improvement of approximately $0.4$ dB over the enumerated algorithm with $V=4$ ($\mathcal{O}\left(16\right)$), while having lower computational complexity. Consequently, the non-enumerated GPS is deemed more efficient in practical implementations. Additionally, a comparison with OTFS shows that AFDM with GPS ($V=2$) achieves better PAPR performance while having lower modulation complexity, since OTFS employs two-dimensional modulation whose complexity is significantly higher than the one-dimensional modulation in AFDM.

\begin{figure}[h]
	\centering
	\includegraphics[width=5.8cm]{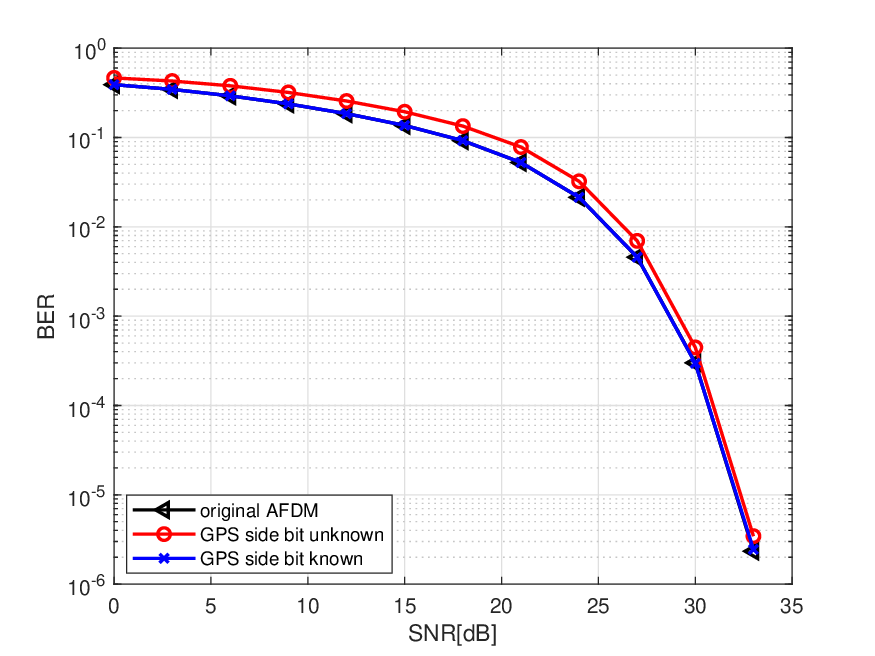}
	\caption{BER performance of 16QAM AFDM with GPS.}
	\label{Fig_5}
\end{figure}

Fig. \ref{Fig_5} presents the bit error rate (BER) performance of AFDM with GPS achieved with additional side information, which is found to be closely comparable to the original AFDM. Targeting a BER performance of $10^{-5}$, the signal-to-noise ratio (SNR) loss resulting from the GPS is about $0.25$ dB. With prior knowledge of the side information, there is no BER performance degradation, confirming that AFDM with GPS preserves the properties of the original AFDM. Additionally, when $N=64$, $V=4$, and $W=2$, the spectral efficiency is about $98.5\%$ of that in original AFDM. To sum up, the results indicate that the GPS exerts a slight impact on the communication performance.

\section{Conclusion}
In this letter, a GPS algorithm is proposed to reduce PAPR in AFDM. We first modify the pre-chirp module of conventional AFDM, ensuring that such an adjustment preserves the inherent characteristics of AFDM. Subsequently, we discuss three aspects in PAPR reduction scheme to achieve the optimal balance between performance and complexity. Finally, simulations are conducted to verify the effectiveness of the proposed GPS. With an acceptable increase in complexity and a marginal degradation in communication performance, AFDM with GPS can achieve a lower PAPR than OTFS, while maintaining a lower modulation complexity.

\ifCLASSOPTIONcaptionsoff
  \newpage
\fi

\bibliographystyle{IEEEtran}
\bibliography{IEEEabrv,mybib}

\end{document}